# Mode-multiplexing deep-strong light-matter coupling

J. Mornhinweg[1,2], L. Diebel[1], M. Halbhuber[1], M. Prager[1], J. Riepl[1],

T. Inzenhofer[1], D. Bougeard[1], R. Huber[1,†], and C. Lange[2,†]

[1]*Department of Physics, University of Regensburg, 93040 Regensburg, Germany*

[2]*Department of Physics, TU Dortmund University, 44227 Dortmund, Germany*

**Dressing quantum states of matter with virtual photons can create exotic effects ranging from vacuum-field modified transport to polaritonic chemistry, and may drive strong squeezing or entanglement of light and matter modes. The established paradigm of cavity quantum electrodynamics focuses on resonant light-matter interaction to maximize the coupling strength $\Omega_R/\omega_c$, defined as the ratio of the vacuum Rabi frequency and the carrier frequency of light. Yet, the finite oscillator strength of a single electronic excitation sets a natural limit to $\Omega_R/\omega_c$. Here, we demonstrate a new regime of record-strong light-matter interaction which exploits the cooperative dipole moments of multiple, highly non-resonant magnetoplasmon modes specifically tailored by our metasurface. This multi-mode coupling creates an ultrabroadband spectrum of over 20 polaritons spanning 6 optical octaves, vacuum ground state populations exceeding 1 virtual excitation quantum for electronic and optical modes, and record coupling strengths equivalent to $\Omega_R/\omega_c = 3.19$. The extreme interaction drives strongly subcycle exchange of vacuum energy between multiple bosonic modes akin to high-order nonlinearities otherwise reserved to strong-field physics, and entangles previously orthogonal electronic excitations solely via vacuum fluctuations of the common cavity mode. This offers avenues towards tailoring phase transitions by coupling otherwise non-interacting modes, merely by shaping the dielectric environment.**

Strong coupling of light and matter is governed by the exchange of energy via virtual excitations, at the Rabi frequency, $\Omega_R$. In the ultrastrong[1–15] ($0.1 \leq \Omega_R/\omega_c < 1$) and deep-strong[7,9,14] coupling (DSC) regime ($\Omega_R/\omega_c > 1$), the vacuum ground state is characterized by an unusually large population of virtual excitations[16,17] with non-classical properties, including significant single or two-mode squeezing[16,17]. This results in spectacular effects of cavity quantum electrodynamics (c-QED)[18] including the vacuum Bloch-Siegert shift[10], polaritonic chemistry[19–21], the creation of photon-bound excitons[22], vacuum-field induced charge transport[13], strong nonlinearities[23], tunnelling[24], and coherent polariton scattering[25]. Landau polaritons[4,5,7,26] of plasmonic THz resonators were the first optical systems to enter the DSC regime[7,9,14] with $\Omega_R/\omega_c = 1.43$ and a large ground state population of 0.37 virtual photons[7]. Moreover, they enable non-adiabatic switching[27] of the extremely squeezed quantum vacuum ground state, which may enable the observation of Unruh-Hawking radiation[28,29] with a further boost of $\Omega_R/\omega_c$. While previous investigations have demonstrated remarkable progress in this direction, they also discovered fundamental barriers given by light-matter decoupling[30] or dissipation[31]. One of the most significant restrictions for increasing $\Omega_R/\omega_c$, however, is the paradigm of resonant light-matter interaction of a single photonic and a single electronic mode, limiting the maximum contributing oscillator strength (Fig. 1a).

Here, we present a conceptually novel approach to DSC which overcomes the limitations of resonant coupling and establishes new record coupling strengths: for sufficiently large spatial overlap of light and matter polarization fields, even electronic excitations that are strongly detuned from the optical mode can substantially boost the vacuum ground state population. We leverage this idea by multiplexing the interaction, exploiting multiple, non-resonant plasmon modes which simultaneously and cooperatively couple with extreme strength to several optical modes of a metallic metasurface (Fig. 1b). Our resonator custom-tailors the plasmon quantum states, while significantly reducing the cavity size as compared to previous approaches, and provides large near-field enhancement. The resulting strongly subcycle transfer of vacuum energy between optical and electronic modes leads to an ultrabroadband spectrum of more than 20 distinct, extremely strongly coupled resonances distributed over up to 6 optical octaves. The vacuum ground state is populated by up to 1.17 virtual photons, which leads to significant squeezing and corresponds to a record coupling strength of $\Omega_R^s/\omega_s = 3.19$ for a hypothetical single pair of light and matter modes, for reference. Surpassing previous values almost twofold, this opens up new possibilities for a range of non-adiabatic c-QED phenomena such as vacuum-field modified transport[13], control of cavity chemistry[19–21], and Unruh-Hawking radiation[28,29]. Moreover, our cavity mediates extremely strong coupling of plasmons that are orthogonal in their bare state - a unique hallmark of the novel regime of deep-strong multi-mode coupling, which may in the future be exploited to drive phase transitions[32] in quantum materials, solely by the vacuum field.

Our resonator design capitalizes on the principles of subwavelength near-field confinement[33] of THz vacuum fields and current distributions in metasurfaces[34], exploiting their great flexibility for designing the spectral shape, resonance frequencies, and near-field structure of optical modes. The structures are specifically optimized for multi-mode coupling and address the two central challenges of our c-QED scenario: designing a broadband spectrum of multiple distinct electronic modes to boost the coupling strength as well as reducing the unit cell size multifold over existing designs. The metasurface is inspired by an inverted layout[35] (Fig. 1c). As a key novelty, we discard the thought pattern of individual, spatially isolated resonators which are generously separated to avoid nearest-neighbour couplings. Instead, we perform finite-element frequency-domain (FEFD) calculations of the current distribution within the metal layer to identify and remove areas of low current density, which are irrelevant to the performance of the resonator. In addition, we exploit the symmetries of the structure to merge adjacent current paths of opposite phase, allowing us to eliminate the affected elements entirely (see Supplementary Information). The resulting metasurface (Fig. 1d) is highly compact and, as a result of the changed topology, can no longer be separated into individual resonators. More precisely, its unit cell is approximately only a quarter as large as compared to conventional designs,[4,7] while a similar near-field enhancement is achieved. Its fundamental LC resonance has a centre frequency of $\nu_{j=1} = 0.52$ THz, while the second, dipolar mode lies at $\nu_{j=2} = 1.95$ THz (Fig. 1e), with $j$ denoting the cavity mode index.

We experimentally evaluate our multi-mode concept by fabricating the gold resonators by electron-beam lithography on top of n-doped GaAs multi-quantum well (QW) heterostructures which host two-dimensional electron gases with a nominal charge carrier concentration of $\rho_{QW} = 1.8 \times 10^{12}$ cm$^{-2}$, and are separated by AlGaAs barrier layers[7] (see Methods). A magnetic bias field $B$ of up to 5.5 T is oriented perpendicularly to the QW plane and Landau-quantizes the electrons, achieving a cyclotron resonance (CR) with a tuneable frequency $\nu_c = eB/2\pi m^*$, where $e$ is the elementary charge and $m^*$ is the electron effective mass.

In the assembled structure, the periodicity of the field localization of the metasurface quantizes the wave vectors of the light field that can couple to the electrons in the QWs (Fig. 2a). As a result, the mode's vacuum field couples to a fan of distinct plasmon resonances[31]. The plasmons feature a frequency of $\omega_P(\boldsymbol{q}_x) = \sqrt{\frac{\rho_{2D} e^2}{2m^* \epsilon_0 \epsilon_{\text{eff}}(\boldsymbol{q})}|\boldsymbol{q}_x|}$ (Refs. [36,37]) (Fig. 2a, red graph), where $\rho_{2D}$ is the total carrier density integrated over all QWs, and $\epsilon_0$ and $\epsilon_{\text{eff}}(\boldsymbol{q})$ are the vacuum permittivity and effective dielectric function, respectively. The plasmon wave vector $\boldsymbol{q}_x(\alpha) = \frac{2\pi}{L_x}\alpha$ depends on the unit cell size in $x$-direction, $L_x$ and the mode index, $\alpha \in \mathbb{Z}$ (see Supplementary Information for the full theory). The small value of $L_x = 30\,\mu\text{m}$ favourably increases the energy spacing as compared to previously investigated structures with larger unit cells[7], leading to more distinct resonances. Plasmon excitation is accounted for up to a maximum index $|\alpha_c| = 10$, beyond which the near-field amplitude is negligible. Linear combinations of plasmon pairs $(-\boldsymbol{q}_x, \boldsymbol{q}_x)$ form bright and dark standing-wave modes which hybridize with the CR at $\omega_c = 2\pi\nu_c$, forming $2|\alpha_c| + 1 = 21$ magnetoplasmon (MP) resonances (see Supplemental Information). Each MP with a frequency of $\omega_{\text{MP},\alpha} \propto \sqrt{\omega_{P,\alpha}^2 + \omega_c^2}$ (Ref. [38]) (Fig. 2b) couples to the resonator modes, $j$, with individual vacuum Rabi frequencies $\Omega_{R,j,\alpha}$ depending on the MP amplitude (Fig. 2a, blue bars) and the detuning, $\nu_j - \omega_{\text{MP},\alpha}/2\pi$. For any given value of $\nu_c$, the detuning vanishes only for a single MP mode at a time (Fig. 2b).

The coupling of the multiple plasmon modes to the electromagnetic modes of our resonator results in a set of multiple light-matter coupled modes. Each of these modes is a characteristic superposition of all MP modes and, since the cavity modes are almost orthogonal[15], only a single cavity mode. We thus reuse the cavity mode index $j$ and introduce the index $\beta$ for the resulting 22 magnetoplasmon cavity polaritons (MPPs). These are categorized into a lower polariton, $\text{LP}_{j,(\beta=0)}$, and, as a novelty of multi-mode coupling, several upper polaritons, $\text{UP}_{j,1\leq\beta\leq\alpha_c+1}$. The presence of only one LP mode, but several UP modes is the fingerprint of cooperative coupling of all independent MP modes to one common cavity mode. Dark modes ($\beta < 0$) are included in our model but not further discussed here (see Supplemental Information). As shown in the simplified example in Fig. 2c, all MPP frequencies rise monotonically when $\nu_c$ is increased from zero. The LP$_{j=1}$ asymptotically emerges from below the CR (dashed line) near $\nu_c \approx 0$ and converges towards the cavity frequency for large $\nu_c$. The UP$_{j=1,\beta}$ modes, for $\nu_c = 0$, start with a finite frequency, which is defined by the plasma frequencies of the constituent MPs as well as the diamagnetic shift caused by the photon field[16]. All UP modes bend upwards with increasing $\nu_c$ and asymptotically converge towards the CR.

In a first experimental campaign, we investigate this new regime of multi-mode coupling for four samples containing 1, 3, 6, and 12 QWs, respectively, by THz time-domain magneto-spectroscopy at cryogenic temperatures as a function of $\nu_c$ (see Methods). For the single-QW structure (Fig. 2d), the single LP$_1$ mode is located below the CR for vanishing $\nu_c$ and approaches the frequency of the bare first cavity mode at 0.5 THz, as $\nu_c$ is increased. For $\nu_c = 0$, the UP$_{1,\beta}$ are observed as a dense fan of partially overlapping modes distributed from ~0 THz upwards. The most strongly visible UP mode, highlighted by the uppermost semi-transparent white curve, is highest in frequency with $\nu = 0.78$ THz. Increasing $\nu_c$, the entire UP mode structure curves upwards in frequency as discussed for Fig. 2c and occupies the increasingly smaller spectral bandwidth between the CR and the highest-energy UP mode. Similarly, the LP$_2$ mode related to the second cavity mode (lowermost dotted curve) branches off below the CR as $\nu_c$ is increased and reaches $\nu = 1.65$ THz at $\nu_c = 1.9$ THz. The UP$_{2,\beta}$ ensemble (upper dotted curves) is dominated by a single spectral feature centred near $\nu = 2.0$ THz for $\nu_c = 0$ THz while slightly shifting upwards with increasing $\nu_c$.

For $n_{QW} = 3$, the overall electronic dipole moment is boosted and the frequency spacing $\omega_P(\alpha)$ of the uncoupled MPs and, as a result, that of the coupled MPPs is increased (Fig. 2e). For $\nu_c = 0$ THz, the UP$_{1,\beta}$ extend from ~0 to 1.0 THz and form clearly separated resonances, evidencing the multi-mode character of the interaction. The increased coupling likewise shifts the frequencies of the UP$_{2,\beta}$ modes further up while lowering the frequencies of the LP$_j$ modes. These trends continue for $n_{QW} = 6$ (Fig. 2f), where for $\nu_c = 0$ THz, the MPP fan manifests in well-separated UP$_{1,\beta}$ modes which extend up to 1.65 THz, whereas the $UP_{2,\beta}$ reach up to 2.0 THz. Yet, owing to the coupling of all MP to the same common cavity mode, only one LP forms for each cavity mode. Implementing $n_{QW} = 12$ (Fig. 2g), we observe distinct $UP_{1,\beta}$ and $UP_{2,\beta}$ branches at frequencies reaching 2.5 THz and 3.2 THz, several of which are simultaneously ultrastrongly coupled with centre frequencies reaching up to ≈5 times the bare cavity frequency, $\nu_{j=1} = 0.52$ THz.

Such extremely strong, off-resonant coupling of multiple light and matter modes with highly disparate frequencies represents a novel setting of c-QED in which strongly subcycle exchange of vacuum energy between the coupled modes is expected. For a quantitative understanding, we developed a multi-mode theory of DSC based on the following Hopfield Hamiltonian[26,39]:

$$\hat{\mathcal{H}} = \sum_j \hbar\omega_j \hat{a}_j^\dagger \hat{a}_j + \sum_\alpha \hbar\omega_{MP}(\alpha) \hat{b}_\alpha^\dagger \hat{b}_\alpha + \sum_{\alpha,j} \hbar\Omega_{R,j,\alpha}(\hat{a}_j^\dagger + \hat{a}_j)(\hat{b}_\alpha^\dagger + \hat{b}_\alpha)$$
$$+ \sum_{\alpha,j} \frac{\hbar\Omega_{R,j,\alpha}^2}{\omega_{MP}(\alpha)}(\hat{a}_j^\dagger + \hat{a}_j)^2 + \hat{\mathcal{H}}_{ext}.$$
(1)

The first two terms describe the bare cavity and MP resonances with frequencies $\omega_j = 2\pi\nu_j$ and $\omega_{MP}(\alpha)$ and bosonic annihilation operators $\hat{a}_j$ and $\hat{b}_\alpha$, respectively. The third term describes their mutual coupling, quantified by a set of vacuum Rabi frequencies, $\Omega_{R,j,\alpha}$, and the fourth term represents the diamagnetic cavity blue shift while $\hat{\mathcal{H}}_{ext}$ implements coupling to external fields.

The individual vacuum Rabi frequencies, $\Omega_{R,j,\alpha}$, are chosen proportional to the near-field amplitude component at the corresponding MP wave vector obtained by Fourier transform of the cavity field, and are scaled by a common factor adjusted for best agreement with the experiment. We include the decay of the near-field amplitude in growth direction, thereby accounting for the coupling to each QW individually (Supplementary Information). Solving Heisenberg's equations of motion for our Hamiltonian (see Methods) unveils the temporal evolution and, by Fourier transform, the spectral shape of the coupled modes. The theory accurately reproduces the measured spectra (Fig. 2h-k) regarding the line widths and oscillator strengths of all modes over the complete experimentally accessible spectral range from <0.1 THz to 6 THz, with quantitative precision. With multiple matter modes off-resonantly coupled to multiple light modes with extreme strengths, the notion of the anti-crossing point that maximizes the coupling strength of a single pair of modes becomes irrelevant. More specifically, when $\Omega_{R,j,\alpha}/\omega_{MP}(\alpha) \approx 1$ is simultaneously reached for several MP modes $\alpha$, the field fluctuations of off-resonant modes influence the vacuum ground state $|G\rangle$ only slightly less than resonant ones (see Supplementary Information). In this setting, the number of virtual photons $\langle N_j \rangle = \langle G|\hat{a}_j^\dagger \hat{a}_j|G\rangle$ of each photonic mode $j$ is the appropriate figure of merit of DSC. Moreover, the relaxed resonance criterion and the coupling to common cavity modes allows for the total vacuum photon population $\sum_j \langle N_j \rangle$ to be increased almost arbitrarily by adding electronic oscillator strength within a spectral window up to several octaves wide.

For our 1-QW structure, we determine $\langle N_1 \rangle = 0.07$ and $\langle N_2 \rangle = 4 \times 10^{-3}$. For comparison with previous demonstrations, the hypothetical coupling strength of a single pair of resonant light and matter modes that yields an equivalent vacuum photon number, $\eta_j = \Omega_{R,j}^s/\omega_j$, amounts to $\eta_1 = 0.55$ and $\eta_2 = 0.13$ for the first two cavity modes. The calculation for the 3-QW structure results in an increased number of virtual photons and effective coupling strength for the cavity modes (Tab. T1). The model also reproduces the transition from the merged ensemble of $UP_{1,\beta}$ modes (Fig. 2h) to clearly distinguishable individual MPP resonances (Fig. 2i-k). In addition, the calculation replicates the data for $n_{QW} = 6$ (Tab. T1). For $n_{QW} = 12$ we reach $\langle N_1 \rangle = 0.76$, $\langle N_2 \rangle = 0.08$ and $\eta_1 = 2.32, \eta_2 = 0.60$. While these values already exceed previous records significantly[7,14], we push the limits of our approach even further with two additional structures. Employing up to 48 QWs, we occupy almost the entire available cavity mode volume and further boost the electronic oscillator strength. The resulting deep-strongly coupled polariton modes extend over an even wider frequency range and create a highly structured spectrum (Fig. 3a,c).

Our calculations (Fig. 3b,d) confirm yet higher vacuum photon populations and coupling strengths for $n_{QW} = 24$ (Tab. T1), achieving $\langle N_1 \rangle = 1.00$, $\langle N_2 \rangle = 0.17$ as well as $\eta_1 = 2.83, \eta_2 = 0.88$ for $n_{QW} = 48$. Here, for the first time, the vacuum photon population of a single coupled optical mode reaches unity, while the combined ground state population of both modes, $\langle N \rangle = \langle N_1 \rangle + \langle N_2 \rangle = 1.17$, comfortably surpasses it. The latter value would be found for an effective coupling strength of $\Omega_R^s/\omega = 3.19$ – exceeding existing structures with $\eta = 1.43$ (Ref. [7]) and $\eta = 1.83$ (Ref. [14]) by almost a factor of 2. The strong back-action of the cavity vacuum fields moreover leads to a combined population of the MPs by 1.06 virtual excitations, and results in strong mixing of the formerly orthogonal matter modes. This perspective holds out the prospect of wide-ranging control of transport[13], chemical reactions[19–21] or phase transitions[32], merely by vacuum fluctuations.

The unconventional nature of this extremely strong multi-mode mixing is especially evident when the subcycle dynamics are investigated directly in the time-domain. To this end, we analyse the polarization dynamics of our structure by the transmitted THz field of the 48-QW sample. At $\nu_c = 0.52$ THz, the experimental waveform consists of a pronounced initial cycle at a delay time of $t = 0$ ps (Fig. 4a). From $t = 0.5$ ps onwards, rapid trailing oscillations are observed which exhibit beating patterns (indicated by arrows) resulting from the rapid energy exchange between multiple light and matter modes. The spectrum has a global maximum located near 2 THz and several small, adjacent local maxima resulting from individual MPP modes (Fig. 4a, inset). Calculating the internal dynamics by our time-domain theory we find that the electric field of the first cavity mode (Fig. 4b, black curve) oscillates on time scales much faster than the cycle duration of the bare mode, $(\nu_1)^{-1}$, (Fig. 4b, grey-shaded area). Its spectrum features 8 distinct local maxima of comparable amplitude located between 0.2 THz, where the LP$_1$ is situated, and up to 4.5 THz (Fig. 4b, inset). Exceeding the frequency $\nu_1$ of the uncoupled cavity by almost an order of magnitude, these features directly result from subcycle energy transfer driven by extreme light-matter coupling strengths, $\Omega_{R,j=1,\alpha}$. Further non-vanishing contributions extend to as low as 0.05 THz and up to 6 THz for a total spectral bandwidth of more than 6 optical octaves. Similarly, the polarization of the first MP mode ($\nu_{MP,\alpha=1} = 1.2$ THz) is strongly structured by multiple oscillations (Fig. 4c). Its spectrum is characterized by four main contributions from 0.18 to 2.59 THz and further, less pronounced components at higher frequencies (Fig. 4c, inset). Moreover, since the large coupling strengths $\Omega_{R,j=1,\alpha}$ to the same cavity mode cause all MP modes to strongly influence each other, the dominant frequencies in their polarization reflect the spectral signatures of all MPPs simultaneously – a unique hallmark of multi-mode non-resonant DSC and the strong mixing of all MPs at the same time (see Supplementary Information). In comparison, the dynamics of the cavity field of a single pair of light-matter coupled modes with $\Omega_R^s/\omega_s = 3.19$ is much less structured (Fig. 4d), and its spectrum contains only the lower and upper polariton resonances (Fig. 4d, inset).

An even stronger contrast between the settings of single and multi-mode coupling is evident from the dynamics of the energy redistribution: Whereas energy exchange of a single pair of modes progresses periodically at $\Omega_R^s$ (Fig. 4e-g, dotted curves), here, the large number of participating modes leads to irregularly structured dynamics for the cavity energy, each of the MP modes, and the energy stored in the coupling[27] (Fig. 4e-g, solid curves). Moreover, the extreme nature of the coupling manifests in its vacuum ground state, where strong squeezing of the photonic mode is observed (Fig. 4h). Further characteristics range from a highly non-classical Fock-state probability distribution (see Supplementary Information) to the emission of correlated photon pairs, expected for non-adiabatic modulation of this exotic vacuum ground state[16].

In conclusion, our work represents a leap forward in c-QED by overcoming the limitations of resonant light-matter coupling with our novel concept of cooperative, multi-mode hybridization, allowing for a significant boost of light-matter coupling strengths. To this end, we designed a maximally compact resonator metasurface that custom-tailors multiple plasmon resonances as well as optical modes to form an ultrabroadband spectrum of Landau cavity polaritons covering 6 optical octaves. This extremely strong coupling results in a highly subcycle vacuum energy exchange and a vacuum ground state hosting a record population of 1.17 virtual photons and 1.06 virtual magnetoplasmon excitations. These figures correspond to a coupling strength of $\Omega_R^s/\omega_s = 3.19$ which exceeds previous records of THz c-QED more than twofold. As our structures permit femtosecond optical switching[27], our tenfold increase of the areal density of virtual excitations will highly benefit the detection of vacuum radiation[29,40]. The extremely strong coupling of multiple electronic modes to the common cavity mode even facilitates hybridization of otherwise orthogonal matter states and can be applied to interactions between a variety of systems such as magnons, phonons or Dirac electrons, including the mixing of entirely different excitations. Combined with the resulting sizeable virtual population of matter and light modes, our concept offers new flexibility and an unprecedented level of control which can be leveraged, for example, in electronic transport[13], light-induced phase transitions[32] or chemical reactions[19–21], by the interaction with vacuum fields.

The authors thank Dieter Schuh and Imke Gronwald for valuable discussions and technical support. We gratefully acknowledge support by the Deutsche Forschungsgemeinschaft through Project IDs 231111959 and 422 31469 5032-SFB1277 (Subproject A01), grants no. LA 3307/1-2, and HU 1598/2, as well as by the European Research Council (ERC) through Future and Emerging Technologies (FET) grant no. 737017 (MIR-BOSE).

J.M., M.H, D.B., R.H. and C.L. conceived the study. M.P., M.H. and D.B. designed, realised, and characterised the semiconductor heterostructures. J.M., M.H. and L.D. modelled the metasurfaces and fabricated the samples with support from T.I., J.R., D.B. and C.L. J.M., M.H., L.D. and J.R. carried out the experiments with support from R.H. and C.L. The theoretical modelling was carried out by J.M., R.H and C.L. D.B., R.H. and C.L. supervised the study. All authors analysed the data and discussed the results. J.M., R.H. and C.L. wrote the manuscript with contributions from all authors.

**Methods**

**Semiconductor heterostructure growth and electron-beam lithography**

Our semiconductor heterostructures were grown by molecular-beam epitaxy on undoped (100)-oriented GaAs substrates which were prepared by growing an epitaxial GaAs layer of a thickness of 50 nm followed by an $Al_{0.3}Ga_{0.7}As$/GaAs superlattice to obtain a defect-free, atomically flat surface. The GaAs quantum well (QW) stacks were embedded in $Al_{0.3}Ga_{0.7}As$ barriers. Si δ-doping layers were placed symmetrically around the individual QWs, enabling control of the carrier density $\rho_{QW}$ of the two-dimensional electron gases (2DEGs). Throughout the manuscript, $\rho_{QW}$ is the density per QW, while $\rho_{2D}$ denotes the combined carrier density of the QW stack, integrated over all QWs along the $z$-direction. The QW stacks were capped by a GaAs layer of a thickness of 30 nm (20 nm for the 12-QW sample) for protection against oxidation.

The THz resonators were fabricated on the surface of the semiconductor structures by electron-beam lithography and wet-chemical processing of polymethylmethacrylat (PMMA) resist followed by thermal vapour-phase deposition of 5 nm of Ti, improving adhesion of the subsequently deposited Au layer of a thickness of 100 nm.

**Wave vector decomposition of resonator near field, and magnetoplasmon modes**

The periodicity of our metasurfaces leads to a discretization of the plasmon wave vectors, $\boldsymbol{q}_x(\alpha) = \frac{2\pi}{L_x}\alpha, \alpha \in \mathbb{Z}$, where $L_x$ is the unit cell size of the structure in $x$-direction, and $\alpha$ is the plasmon mode index. For each $\boldsymbol{q}_x(\alpha)$, the relevant electric field amplitude of the cavity mode is given by the Fourier component $[\mathcal{F}(\mathcal{E}_x)](\boldsymbol{q}_x, \boldsymbol{q}_y = 0)$, with $\mathcal{F}(\mathcal{E}_x)$ denoting the 2D Fourier transform of the electric near-field, $\mathcal{E}_x$, in $x$-direction. $\mathcal{E}_x$ is calculated by the finite-element method for the bare resonator on an undoped substrate. The amplitudes are determined separately for each QW separately. Fig. S3 shows these data for the first two resonator modes, as a function of the wave vector and the depth below the metasurface. For large wave vectors, single-particle excitations become energetically possible and the field amplitude moreover drops sharply, limiting the relevant wave vector spectrum up to a maximum index $|\alpha_c| = 10$.

**Cryogenic THz magneto-spectroscopy**

Femtosecond near-infrared pulses (centre wavelength, 807 nm; pulse energy, 5.5 mJ, pulse duration, 33 fs) from a titanium-sapphire amplifier laser (repetition rate, 3 kHz) were used to generate single-cycle THz pulses by optical rectification and to detect the transmitted waveforms by electro-optic sampling. Depending on the required bandwidth, we employed ⟨110⟩-cut ZnTe crystals of a thickness of 1 mm for our structures with up to 12 QWs, and GaP crystals of a thickness of 200 μm for the 24 and 48-QW structures. A mechanical chopper modulated the THz pulses, allowing for differential detection of the transmitted THz electric field, $E(t)$. The sample was kept at a temperature of 10 K in a magneto-cryostat with a large numerical aperture, enabling magnetic fields up to 5.5 T applied perpendicularly to the sample surface. The recorded electric field was Fourier transformed and referenced to a measurement without a sample in the cryostat to obtain the transmission spectra.

**Sample structure parameters**

For each sample, the doping densities, coupling strengths $\eta_j$ for each cavity mode, and the corresponding vacuum photon numbers, $N_j$, are listed in table T1.

| n_QW | Doping density $\rho_{QW}$ (cm$^{-2}$) | $\langle N_1 \rangle$ | $\langle N_2 \rangle$ | $\langle N \rangle$ | $\eta_1$ | $\eta_2$ |
|---|---|---|---|---|---|---|
| 1 | $1.28 \times 10^{12}$ | 0.07 | 0.004 | 0.07 | 0.55 | 0.13 |
| 3 | $7.9 \times 10^{11}$ | 0.13 | 0.01 | 0.14 | 0.76 | 0.19 |
| 6 | $1.37 \times 10^{12}$ | 0.34 | 0.03 | 0.37 | 1.34 | 0.36 |
| 12 | $1.90 \times 10^{12}$ | 0.76 | 0.08 | 0.85 | 2.32 | 0.60 |
| 24 | $2.30 \times 10^{12}$ | 0.99 | 0.16 | 1.14 | 2.80 | 0.85 |
| 48 | $1.80 \times 10^{12}$ | 1.00 | 0.17 | 1.17 | 2.83 | 0.88 |

**Table T1 | Structural and coupling parameters of the 6 QW structures**. $\langle N_1 \rangle$, $\langle N_2 \rangle$, $\langle N \rangle$: vacuum photon numbers for the first and second resonator modes, and for both combined, respectively. $\eta_1$ and $\eta_2$ denote the hypothetical coupling strengths of, in each case, a single pair of light and matter modes with a vacuum photon population of $\langle N_1 \rangle$ or $\langle N_2 \rangle$, respectively, for comparison.

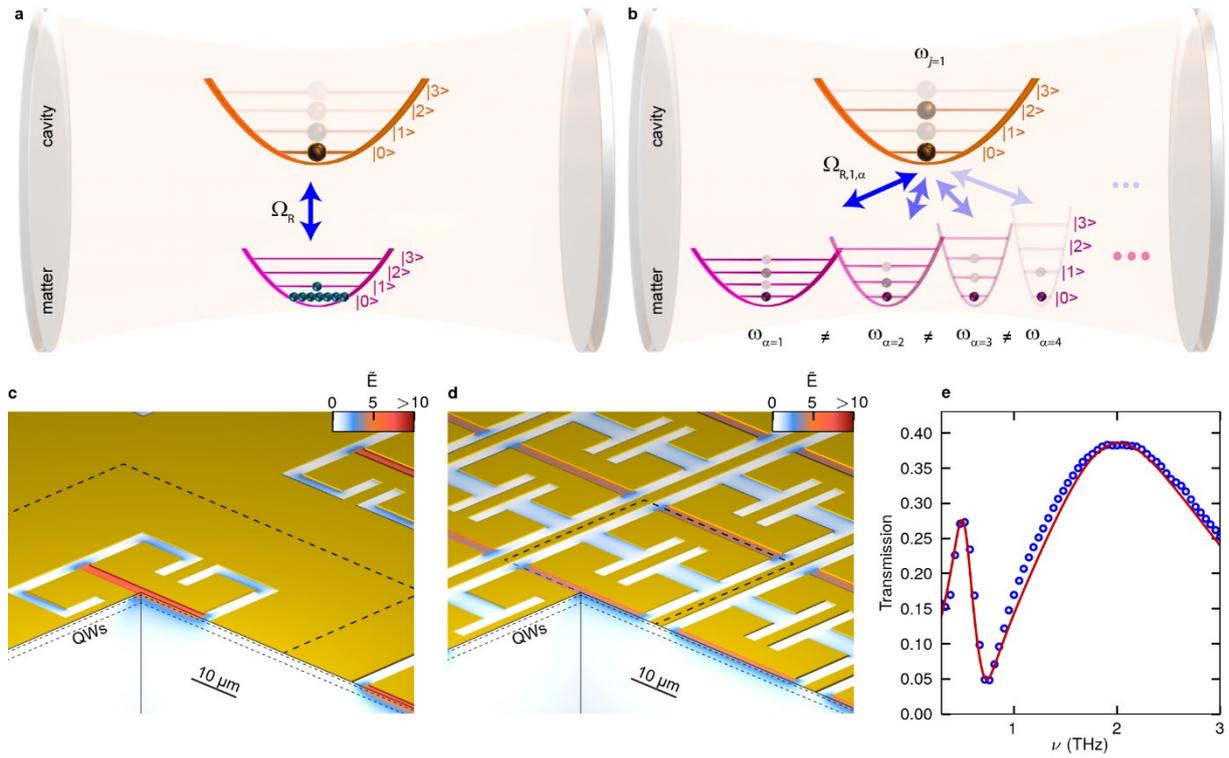

**Figure 1 | Multi-mode light-matter coupling and ultracompact metasurface. a**, Illustration of resonant ultrastrong coupling of a single cavity mode (upper parabola) to a single matter excitation (bottom parabola) with a vacuum Rabi frequency $\Omega_R$. The weak population by virtual excitations in the vacuum ground state is indicated by semi-transparent spheres. **b**, Illustration of deep-strong coupling of one light mode (upper parabola) to multiple matter excitations (bottom parabolas) with vacuum Rabi frequencies $\Omega_{R,j,\alpha}$ under off-resonant conditions. Owing to the extremely large light-matter coupling, a significant number of virtual excitations are present. **c**, Three-dimensional cut-away illustration of a conventional metasurface structure (gold shape) and its electric near-field distribution $\tilde{E}$, for the fundamental LC mode at $\nu_1 = 0.52$ THz. The unit cell (size: 60 μm × 60 μm) is indicated by the dashed line. QW: quantum well stack. **d**, Highly compacted metasurface with a unit cell size of 30 μm × 32.5 μm. **e**, Measured THz transmission of the bare resonator array (blue circles), and calculation (red curve).

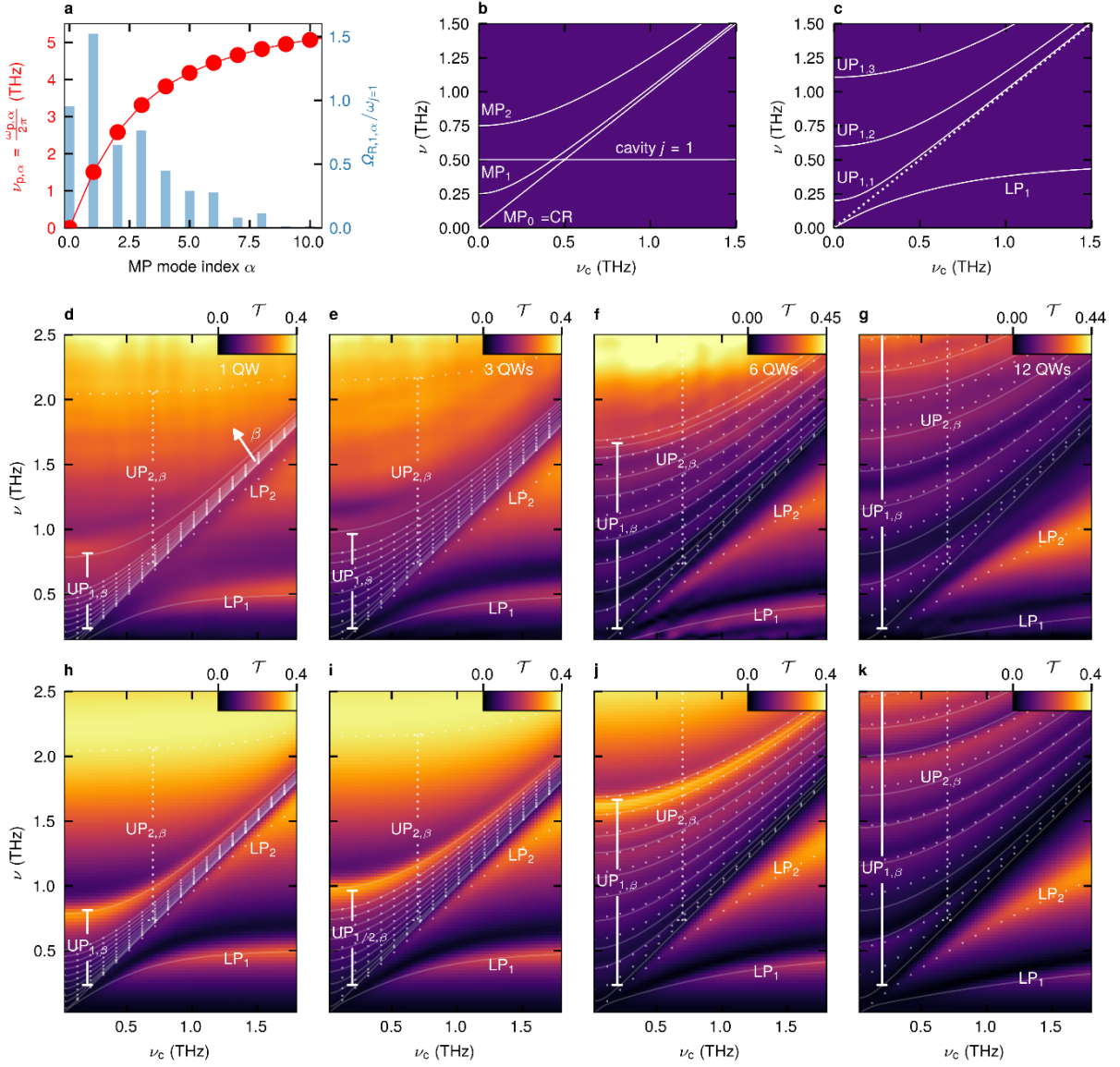

**Figure 2 | Deep-strong light-matter coupling. a**, Plasmon frequency $\nu_{p,\alpha}$ and coupling strength $\Omega_{R,1,\alpha}/\omega_{j=1}$ for each MP mode n in the case of the 48 QW sample. **b,c,** Illustration of multi-mode coupling of one cavity mode to several matter modes, $MP_0 = CR$, $MP_1$ and $MP_2$, as a function of the cyclotron frequency, $\nu_c$. **b**, uncoupled modes. **c**, Coupled modes comprising of one lower polariton ($LP_1$) and three upper polaritons ($UP_{1,1}$, $UP_{1,2}$ and $UP_{1,3}$). **d**, THz magneto-transmission as a function of $\nu_c$ of the single-QW structure. The continuous white curves trace the polariton modes obtained from the multi-mode Hopfield model for the first resonator mode (coupling strength: $\eta_1 = 0.55$). The dashed white curves represent the polaritons linked to the higher mode $\nu_2$ ($\eta_2 = 0.13$). **h**, Spectrum obtained from time-domain quantum model and identical polariton frequencies (dashed curves), for comparison. **e,**

Transmission of the 3-QW structure ($\eta_1 = 0.76$) and **i**, simulation. **f**, Transmission of the 6-QW structure ($\eta_1 = 1.34$), and **j**, simulation. **g**, Transmission of the 12-QW structure ($\eta_1 = 2.32$), and **k**, simulation.

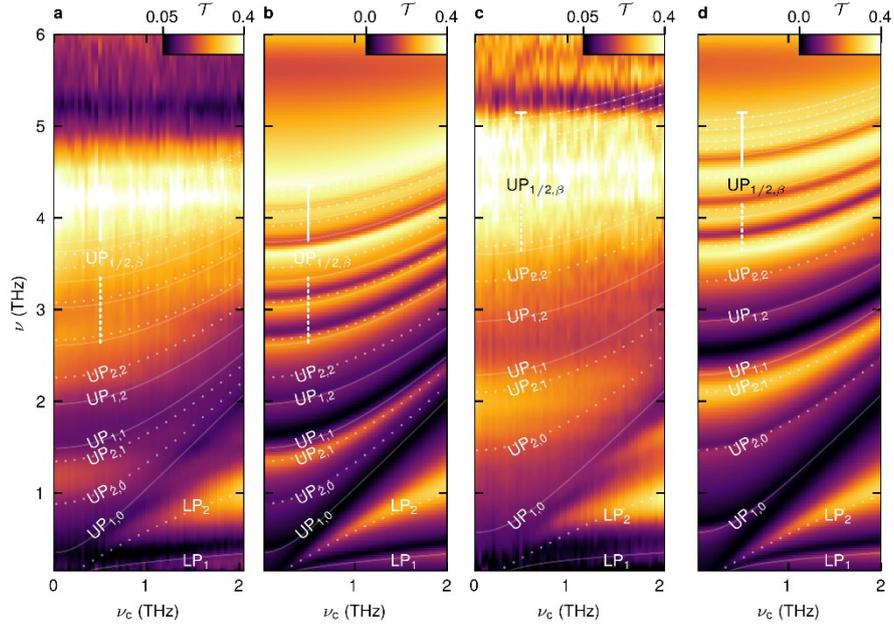

**Figure 3 | Extremely strong, multi-octave light-matter coupling. a**, THz magneto-transmission of the 24-QW sample as a function of $\nu_c$ (see Fig. 2). The extended Hopfield model yields coupling strengths of $\eta_1 = 2.80$ and $\eta_2 = 0.85$ for first and second resonator mode, respectively. Calculated polariton frequencies (solid & dashed curves) with distinct resonances marked. **b**, Calculated transmission and polariton frequencies. **c**, Transmission of the 48-QW structure. Coupling strengths: $\eta_1 = 2.83$, $\eta_2 = 0.88$. **d**, Calculated transmission and identical polariton frequencies.

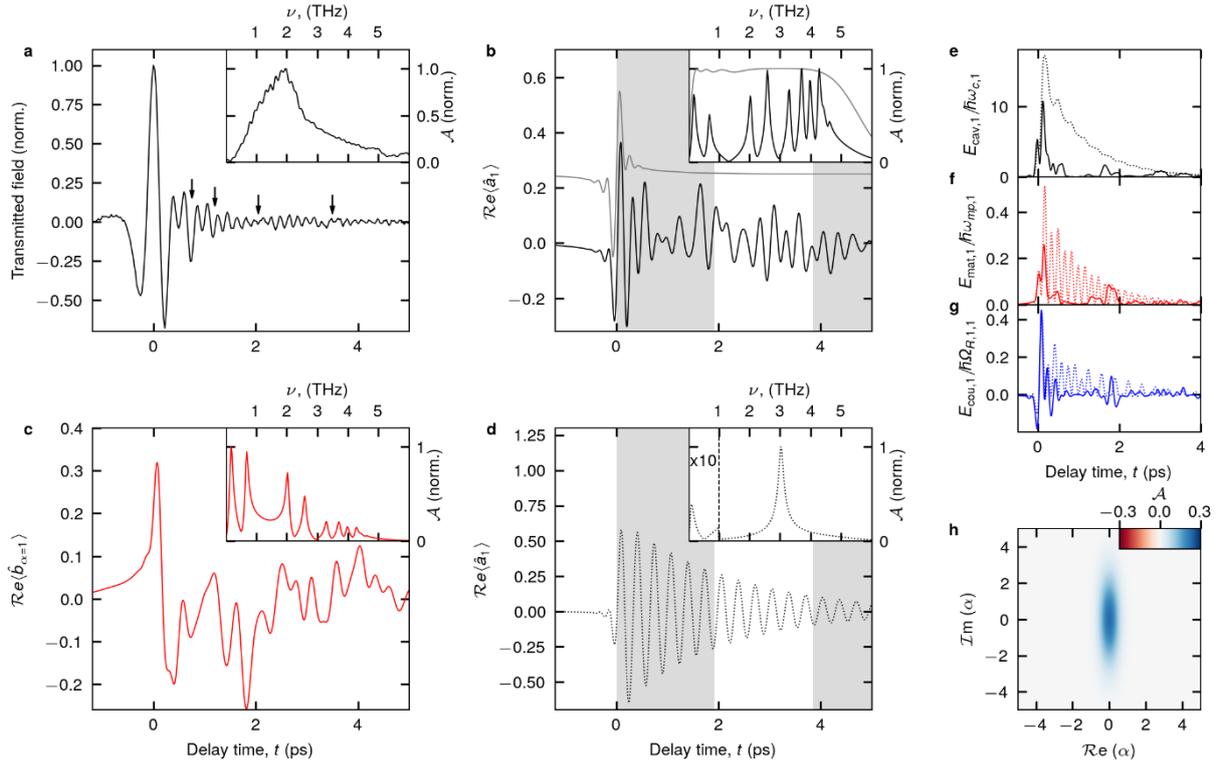

**Figure 4 | Dynamics and squeezing of extremely strong light-matter coupling. a**, Transmitted THz field of the 48-QW sample ($\eta_1 = 2.83$) at $\nu_c = 0.52$ THz (black curve). Inset: Spectral amplitude of the THz field. **b**, Calculated expectation value for the population of the first cavity mode $Re\langle \hat{a}_{j=1} \rangle$ after excitation (black curve) by a broadband pulse (grey curve). Inset: Corresponding spectra. The shading marks one oscillation period of the bare cavity mode. **c**, Calculated expectation value of the polarization of the first MP mode, $Re\langle \hat{b}_{\alpha=1} \rangle$, and spectrum (inset). **d**, $Re\langle \hat{a}_{j=1} \rangle$ for the same coupling strength as in **b**, yet only for a single pair of light and matter modes. Inset: Corresponding spectrum. **e**, Energy of the first cavity mode for the full calculation (solid curve) and the single-mode reference (dashed curve) as in panel **d**. **f**, Energies of the first MP mode (solid curve) and CR (dashed curve) for the two cases. **g**, Corresponding coupling energies between cavity mode and MP mode (solid curve) or CR (dashed curve). **h**, Wigner-function representation for the photonic state of a deep-strongly coupled system with $\Omega_R/\omega_c = 3.0$.

# Mode-multiplexing deep-strong light-matter coupling

# Supplementary Information

J. Mornhinweg[1,2], L. Diebel[1], M. Halbhuber[1], M. Prager[1], J. Riepl[1],

T. Inzenhofer[1], D. Bougeard[1], R. Huber[1,†], and C. Lange[2,†]

[1]*Department of Physics, University of Regensburg, 93040 Regensburg, Germany*

[2]*Department of Physics, TU Dortmund University, 44227 Dortmund, Germany*

# Table of Contents

# 1. Ultracompact metasurface design

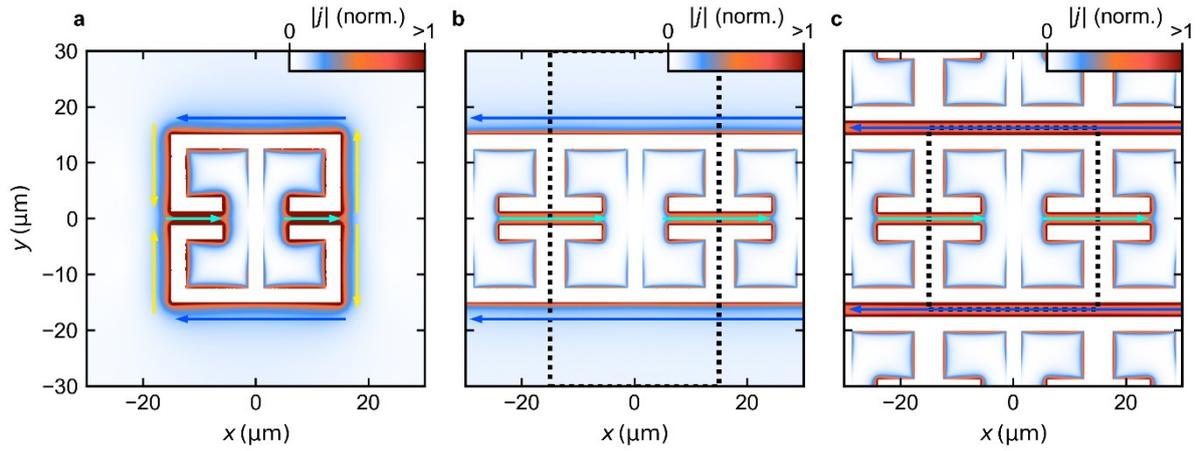

**Fig. S1 | Ultracompact resonator design and current distributions. a**, Resonator layout featuring a conventional, large unit cell. Arrows: Current flow of the fundamental optical mode. **b**, Resonator design compacted in $x$-direction, and current flow. The unit cell is indicated by the dashed rectangle. **c**, Resonator design, compacted in $x$ and $y$-direction, and current flow.

The design of our gold metasurfaces aims at a maximally compact resonator geometry while providing spectrally well-separated optical modes with low linewidths, high near-field enhancement factors as well as low mode volumes. To this end, we have revisited some of the established design principles for plasmonic resonator structures. As detailed in the following, our new approach enables excess spacing between adjacent resonators to be minimized or completely discarded, increasing the areal resonator density and thus the total oscillator strength of the metasurface by a factor of ~4 as compared to the original design.

The inverted resonator structures are based on a square of 30 μm outer extension and twin current paths feeding a central capacitive gap of a width of 2.5 μm. The unit cell is twice as large as the outer dimensions of the resonator. In the frequency window of interest, the structure supports five optical modes with centre frequencies of 0.52, 1.95, 3.75, 4.6 and 6 THz (see main text, Table T1 and Fig. S2). The fundamental mode at 0.52 THz is excited by THz radiation linearly polarized in x direction, leading to an oscillating current flow along the paths indicated in Fig. S1a, whereby the current lags behind the driving field by a phase of $\pi/2$. As the field sweeps charges out of the right inner metal plate (turquoise right arrow), symmetry dictates that a current identical in direction and magnitude drives charges into the corresponding mirror plate on the left side of the structure (left turquoise arrow). Following Kirchhoff's laws, the loop is closed by currents passing through the outer metal plane symmetrically, in $y$ (yellow arrows) and $x$-direction (blue arrows). As a first step, we exploit the anti-symmetry of the $y$-oriented currents along the left and right edges of the outer metal plane. As the spacing of adjacent resonators in x-direction, $s_x$, is reduced, their near-field currents increasingly overlap. Owing to the opposite orientation of the currents marked by yellow arrows, they locally cancel each other. At the same time, the current flow into and out of the internal resonator area (turquoise arrows) remains unaffected since the role of the yellow currents is increasingly taken over by the current flows out of and into the internal resonator area of the nearest resonator neighbour. As a result, all relevant properties of the optical mode remain unaffected. Since this argument holds for any value of $s_x$, we chose $s_x = 0$, for which the metal area separating the resonators in x-direction vanishes completely. Here too, the currents exiting the inner metal plates of one resonator seamlessly continue to flow, now feeding the opposite metal plate of the adjacent resonator (Fig. S1b, turquoise arrows) instead of the now absent $y$-polarized current paths (Fig. S1a, yellow arrows). Likewise, the $x$-polarized currents flowing along the outer perimeter of the structure are now connected between adjacent resonators (blue arrows) which also renders the $y$-polarized paths unnecessary.

We next consider a reduction of the spacing of the structures in $y$-direction, $s_y$. Since the outer currents of adjacent unit cells in $y$-direction carry the same phase, the cancellation exploited above cannot be applied here. However, since the resonance condition for the LC mode favours the shortest overall current paths, the currents are generally concentrated very close to the edges of the metallized layer (Fig. S1), from where they fall off within $\tilde{\rho}_y \approx 3$ µm in the direction normal to the edge, where $\tilde{\rho}_y$ is the characteristic decay length, in $y$-direction. As a consequence, $s_y$ may be significantly reduced as long as the remaining metal stripline is of sufficient width to allow the $x$-polarized current to continue to flow. In the final structure, we chose $s_y = 2.5$ µm $\approx \tilde{\rho}_y$, leading to a reduction of the area of the unit cell by a factor close to 4. While the near-field distribution of this structure is virtually identical to that of the original one, the current distribution is markedly different. As Fig. S1 shows, the structure now exhibits two mainly $x$-polarized subcurrents which sweep charges between the adjacent inner metal plates and along the striplines, respectively, and are offset by a phase of $\pi$. Finally, since the arguments of symmetry and current localization apply to all resonator modes equally, the spectrum of the compacted structure does not differ significantly from the spectrum of the separated structures across the entire spectral range, apart from coupling to surface plasmons [1] and the overall increase in transmission due to the unit cell reduction (Fig. S2).

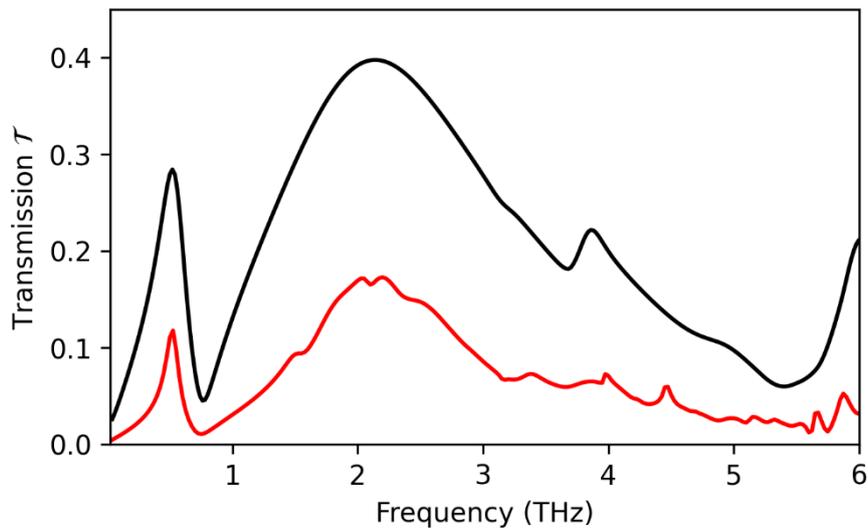

**Figure S2 | Transmission for the bare resonators.** Transmission spectrum for excitation in x-direction of the resonator array featuring conventional, large unit cell (60 μm x 60 μm, red curve) and of the compacted structure (black curve).

## 2. Magnetoplasmon modes in two-dimensional electron gases

In the two-dimensional electron gas (2DEG) hosted in our quantum wells (QWs), in absence of a magnetic field, plasma excitations obey the dispersion relation

$$\omega_{\text{plasma}}(\boldsymbol{q}) = \sqrt{\frac{\rho_{\text{2D}} e^2}{2m^* \epsilon_0 \epsilon_{\text{eff}}(\boldsymbol{q})} |\boldsymbol{q}|}, \tag{1}$$

where $\boldsymbol{q}$ denotes the in-plane wavevector, $\rho_{\text{2D}}$ is the two-dimensional (2D) electron density, $m^*$ is the effective electron mass and $\epsilon_{\text{eff}}(\boldsymbol{q})$ represents the effective dielectric constant for the electron gas [2,3]. Applying a static magnetic bias field oriented perpendicularly to the plane of the 2DEG introduces Landau quantization and the plasmon excitations hybridize with the cyclotron resonance, $\omega_c$, resulting in the formation of magnetoplasmons (MPs) [4] with a frequency of

$$\omega_{\text{MP}} = \sqrt{\omega_c^2 + \omega_{\text{plasma}}^2}. \tag{2}$$

The effective dielectric constant $\epsilon_{\text{eff}}(\boldsymbol{q})$ for the MPs is set by the dielectric environment of the electron gas in the layers enclosing the 2DEG. In our structures, we consider the dielectric constant of the GaAs substrate below the QWs, $\epsilon_{\text{sub}} = 12.9$, the capping layer of a thickness of $d$ and dielectric constant $\epsilon_{\text{barrier}}$ on top of the QWs, as well as the dielectric function of the top interface to either vacuum, or the gold parts of the resonator structure. The latter two possibilities have been analysed previously [5,3], resulting in corresponding effective dielectric functions, $\epsilon_{\text{ungated}}(\boldsymbol{q})$ for the vacuum interface, and $\epsilon_{\text{gated}}(\boldsymbol{q})$ for a top-metallized 2DEG:

$$\epsilon_{\text{ungated}}(\boldsymbol{q}) = \frac{\epsilon_{\text{sub}}}{2} + \frac{\epsilon_{\text{barrier}}}{2} \times \frac{1 + \epsilon_{\text{barrier}} \tanh(|\boldsymbol{q}|d)}{\epsilon_{\text{barrier}} + \tanh(|\boldsymbol{q}|d)} \tag{3}$$

$$\epsilon_{\text{gated}}(\boldsymbol{q}) = \frac{\epsilon_{\text{sub}} + \epsilon_{\text{barrier}} \coth(|\boldsymbol{q}|d)}{2}. \tag{4}$$

Depending on the situation, either of the two functions assumes the role of $\epsilon_{\text{eff}}(\boldsymbol{q})$ in Eq. 1. For more complex, laterally structured samples such as planar metal resonators or gratings, an averaged, effective dielectric function can be constructed, proportionally factoring in the two situations according to a factor $\delta$ describing the relative metal coverage of the surface:

$$\epsilon_{\text{eff,mix}}(\boldsymbol{q}) = \delta\epsilon_{\text{gated}}(\boldsymbol{q}) + (1-\delta)\epsilon_{\text{ungated}}(\boldsymbol{q}). \tag{5}$$

Moreover, the multi-QW stack consists of several layers with varying dielectric properties, requiring additional averaging of the effective dielectric function along the growth direction. As was shown previously, the response of the densely packed QWs is well approximated by an effective-medium approach [6]. In addition, as the QW stack exhibits a finite extension in growth direction, the charge carriers in the QWs are not an ideal 2D electron gas. More precisely, with increasing QW thickness, their plasma frequency approaches the 3D plasma frequency as an asymptotical upper limit for large wave vectors, whereas the dispersion of an ideal 2D plasma has no such upper bound. To account for this effect, we must include a correction which depends on the QW stack thickness $t$. The effective dielectric function $\epsilon_{\text{eff}}$ then is given by [7]:

$$\epsilon_{\text{eff}} = \epsilon_{\text{eff,mix}} + \frac{\epsilon_{\text{sub}}|\boldsymbol{q}|t}{2}. \tag{6}$$

This description allows for a MP mode dispersion approaching the 3D plasma frequency asymptotically for $|\boldsymbol{q}| \to \infty$. However, plasmon and magnetoplasmon excitations are limited in frequency and wave vector by Landau damping which becomes effective when single-particle excitations become relevant [8,4]. An upper bound for the frequency of these single-particle excitations is given by $\nu < v_F q_x$, with the Fermi velocity $v_F = \frac{\hbar\sqrt{2\pi\rho_{2D}}}{m^*}$.

The periodicity of our metasurfaces implies a discretization of the plasmon wave vectors that can be excited by the cavity modes. The corresponding condition for the in-plane wave vectors $\boldsymbol{q}_x$ is

$$\boldsymbol{q}_x(\alpha) = \frac{2\pi}{L_x}\alpha. \tag{7}$$

Here, $L_x$ denotes the unit cell size of the structure in $x$-direction, and $\alpha \in \mathbb{Z}$ is the plasmon mode index. Linear combinations of plasmon waves with wave vectors $-\boldsymbol{q}_x$ and $\boldsymbol{q}_x$ form bright and dark standing waves $\Psi_b \propto \exp(-i\boldsymbol{q}_x x) + \exp(i\boldsymbol{q}_x x)$ and $\Psi_d \propto \exp(-i\boldsymbol{q}_x x) - \exp(i\boldsymbol{q}_x x)$, respectively. Since only the bright modes couple to the cavity modes, the dark modes are not further considered. We verify this approach for our complex resonator geometry by calculating the 2D Fourier transform of the $x$-polarized electric near-field component $\mathcal{E}_x$ of the first two cavity modes (Fig. S3a,b) along the $x$-direction. The field $\mathcal{E}_x$ was calculated by the finite-element (FEM) method (see chapter 4) [6] and evaluated within $xy$-oriented planes that lie within the QW stack, whereby the doping concentration of the QWs was set to zero to obtain the modes of the empty cavity. This choice represents the field component most relevant for light-matter coupling, whereas we neglect the $\mathcal{E}_y$-polarized components since they are significantly weaker in most areas of the structure. The resulting amplitudes, integrated within the respective plane, are shown in Fig. S3c,d as a function of the wave vector and the depth below the metasurface. Owing to the spectrum of single-particle excitations (see previous paragraph) and the fact that the field amplitude decays for larger wave vectors, in particular within planes more distant to the metasurface, we limit the magnetoplasmon mode index $\alpha$ to a maximum of $|\alpha| \leq \alpha_c = 10$. The data allows us to model the coupling of the light field to each QW separately and to account for the relative amplitude of each magnetoplasmon resonance. The same analysis is also performed for the higher resonator mode at 1.95 THz. The resulting plasmon dispersions, the wavevectors fulfilling the discretisation and the cut-off wavevector is shown in Fig. S4, whereas the relative coupling strengths for the plasmon modes in shown in the main manuscript Fig. 3a.

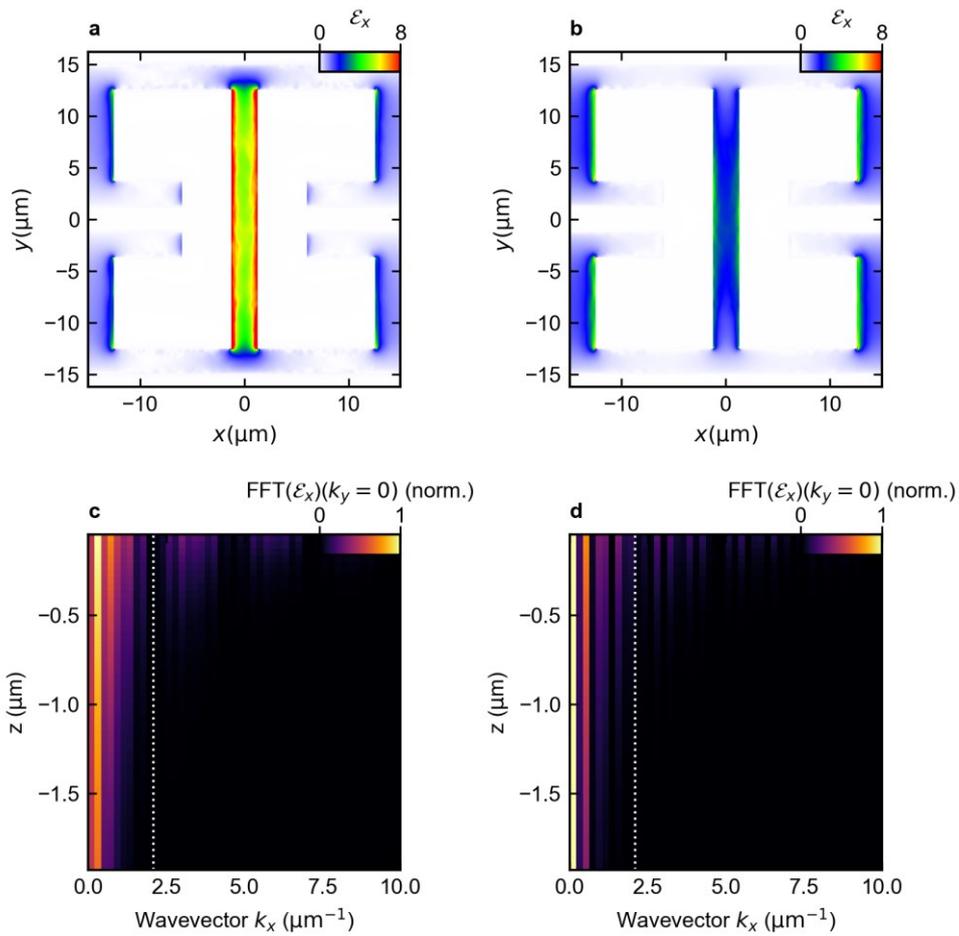

**Figure S3 | Analysis of magneto-plasmon formation.** Electric field component $\mathcal{E}_x$ in the quantum well plane 50 nm below the resonator structure, for **a**, the LC mode ($j = 1$) at 0.52 THz and **b**, the dipolar mode ($j = 2$) at 1.95 THz. **c**, Amplitude components of the Fourier-transformed data of panel **a** as a function of the wave vector $k_x$, for $k_y = 0$, and the depth of the plane below the metasurface, $z$. **d**, Equivalent amplitude components for the data of the dipolar mode in panel **b**.

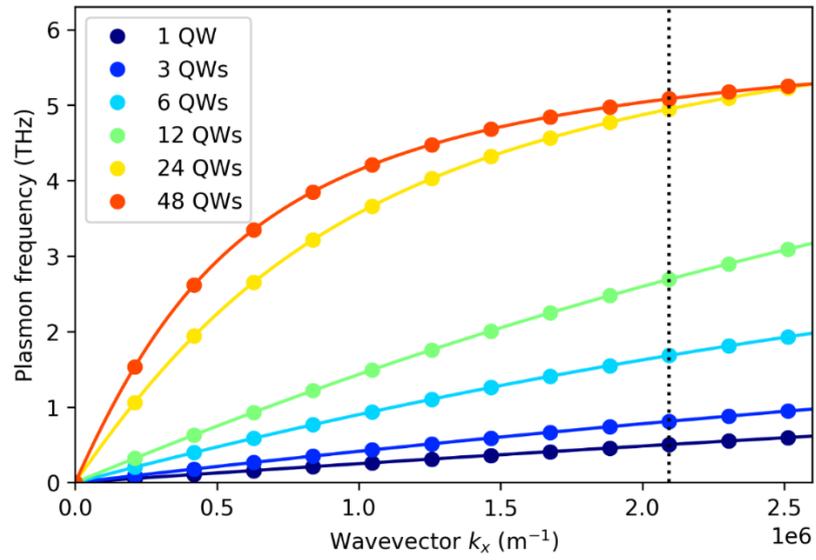

**Figure S4 | Plasmon dispersion for all samples.** The dots mark the plasmon frequencies which fulfil the diffraction condition. The black dashed line indicates the cut-off wavevector up to which plasmon modes were considered in the quantum model.

## 3. Multi-mode quantum model of deep-strong light-matter coupling

We describe the multi-mode light-matter coupling of our structures by a specifically developed mean-field theory treating light-matter interaction in the deep-strong coupling regime, including anti-resonant terms. We start with the Hamiltonian

$$\hat{H} = \hat{H}_{\text{cav}} + \hat{H}_{\text{e}} + \hat{H}_{\text{int}} + \hat{H}_{\text{dia}} + \hat{H}_{\text{ext}}. \tag{8}$$

The cavity contribution is described by

$$\hat{H}_{\text{cav}} = \hbar\omega_{\text{cav}}\hat{a}^\dagger\hat{a}, \tag{9}$$

where $\omega_{\text{cav}}$ is the frequency of the cavity mode, and $\hat{a}$ is the corresponding bosonic annihilation operator. In a first limit of low doping density $\rho$, the frequencies of plasma excitations are much lower than the linewidth of the cyclotron resonance, such that the entire spectrum of excitations approximately coincides with the CR frequency. We implement this situation in the bosonic limit by

$$\hat{H}_{\text{e}} = \hbar\omega_c \hat{b}^\dagger\hat{b}. \tag{10}$$

Here, $\omega_c = 2\pi\nu_c$ denotes the cyclotron frequency and $\hat{b}$ is the annihilation operator of the Landau excitations. Light-matter coupling including anti-resonant interaction terms is introduced by

$$\hat{H}_{\text{int}} = \hbar\Omega_R(\hat{a}^\dagger + \hat{a})(\hat{b}^\dagger + \hat{b}), \tag{11}$$

where $\Omega_R$ is the vacuum Rabi frequency of the cavity mode coupling to the cyclotron resonance. The blue-shift of the cavity modes by diamagnetic interactions is accounted for by

$$\hat{H}_{\text{dia}} = \hbar D(\hat{a} + \hat{a}^\dagger)^2, \tag{12}$$

where $D = \frac{(\Omega_R)^2}{\omega_c}$. Coupling of the cavity modes to the THz far-field is contained in $\hat{H}_{\text{ext}}$.

When the doping density is raised to a significant level, the plasmon frequencies increase such that $\omega_{\text{plasma}}(\boldsymbol{q}(\alpha_c))$ eventually exceeds the linewidth of the cyclotron resonance. As a consequence, the magnetoplasmon modes become the relevant matter excitations in the light-matter coupling Hamiltonian. At the same time, the multiple optical resonances of the resonator structure over several octaves of coupled modes have to be considered. We account for this situation by extending $\hat{H}_{\text{cav}}$ to all cavity modes $j$ with frequencies $\omega_j$, while $\hat{H}_{\text{e}}$ is modified to include all relevant MP resonances up to the cut-off MP index, whereby the magnetoplasmon frequencies are $\omega_{\text{MP}}(\alpha) = \sqrt{\omega_c^2 + \omega_{\text{plasma}}^2(\boldsymbol{q}(\alpha))}$.

Each cavity mode $j$ is coupled to all matter modes simultaneously with an individual Rabi frequency $\Omega_{\text{R},j,\alpha}$. The extended multi-mode Hamiltonian then reads:

$$\hat{\mathcal{H}} = \sum_j \hbar\omega_j \hat{a}_j^\dagger \hat{a}_j + \sum_\alpha \hbar\omega_{\text{MP}}(\alpha)\hat{b}_\alpha^\dagger \hat{b}_\alpha + \sum_{\alpha,j} \hbar\Omega_{\text{R},j,\alpha}(\hat{a}_j^\dagger + \hat{a}_j)(\hat{b}_\alpha^\dagger + \hat{b}_\alpha) \\ + \sum_{\alpha,j} \frac{\hbar\Omega_{\text{R},j,\alpha}^2}{\omega_{\text{MP}}(\alpha)}(\hat{a}_j^\dagger + \hat{a}_j)^2 + \hat{\mathcal{H}}_{\text{ext}}. \quad (13)$$

The individual coupling strengths of the MP modes, $\Omega_{\text{R},j,\alpha}$, are determined by the amplitudes of the associated wave vectors, as obtained by a Fourier transform of the near field of the resonator mode (see previous section). The set of all $\Omega_{\text{R},j,\alpha}$ is then scaled by a single common factor to match the experimental spectra.

We perform a Bogoliubov transformation in order to determine the frequencies of the coupled modes as well as their light-matter composition, which is given by the eigenvalues and eigenvectors of the transform matrix. The resulting normal-mode polariton operators, $\{\hat{p}_{\beta,j}, \hat{p}_{\beta,j}^\dagger\}$, are given by $\hat{p}_{\beta,j} = w_{\beta,j}\hat{a}_j + \sum_\alpha x_{\beta,\alpha}\hat{b}_\alpha + y_{\beta,j}\hat{a}_j^\dagger + \sum_\alpha z_{\beta,\alpha}\hat{b}_\alpha^\dagger$ and generally contain contributions from all magnetoplasmon modes. The Hopfield coefficients $(w_{\beta,j}, x_{\beta,\alpha}, y_{\beta,j}, z_{\beta,\alpha})$ represent the polariton fractions corresponding to the bare cavity and matter modes. The absolute values of the Hopfield coefficients are shown in Fig. S5 for the sample with 48 QWs, for a cyclotron frequency of $\nu_c = 0.52$ THz. As discussed in the main text, each wave vector pair $(-\boldsymbol{q}_x, \boldsymbol{q}_x)$ enables two linear superpositions associated with a dark and a bright magnetoplasmon mode, respectively. For clarity, we enumerated the eigenmodes of the Hamiltonian as follows: dark magnetoplasmon modes which do not couple to the cavity modes have a polariton index $\beta < 0$. The lower polariton mode is attributed to $\beta = 0$. Finally, the bright magnetoplasmon modes which do couple to the cavity field lead to the light-matter hybridized upper polariton modes with indices $\beta > 0$. Following the procedure of Ref. [9], we subsequently calculate the expectation value of the vacuum photon population $\langle N \rangle = \langle G|\hat{a}_j^\dagger \hat{a}_j|G\rangle$ for each cavity mode $j$, whereby $|G\rangle$ denotes the ground state of the coupled system. For our structures, the vacuum photon population $\langle N \rangle$ not only reaches values of up to one entire photon, but the vacuum ground state moreover features exotic properties such as a non-classical Fock state occupation distribution. More precisely, the occupation probability does not diminish monotonically as a function of the photon number but instead shows a higher occupation probability for the excited state $|2\rangle$ as compared to the state $|1\rangle$ (Fig. S6a). In addition, the ground state exhibits strong squeezing as evidenced by the Wigner function for the photonic state (Fig. S6b-d).

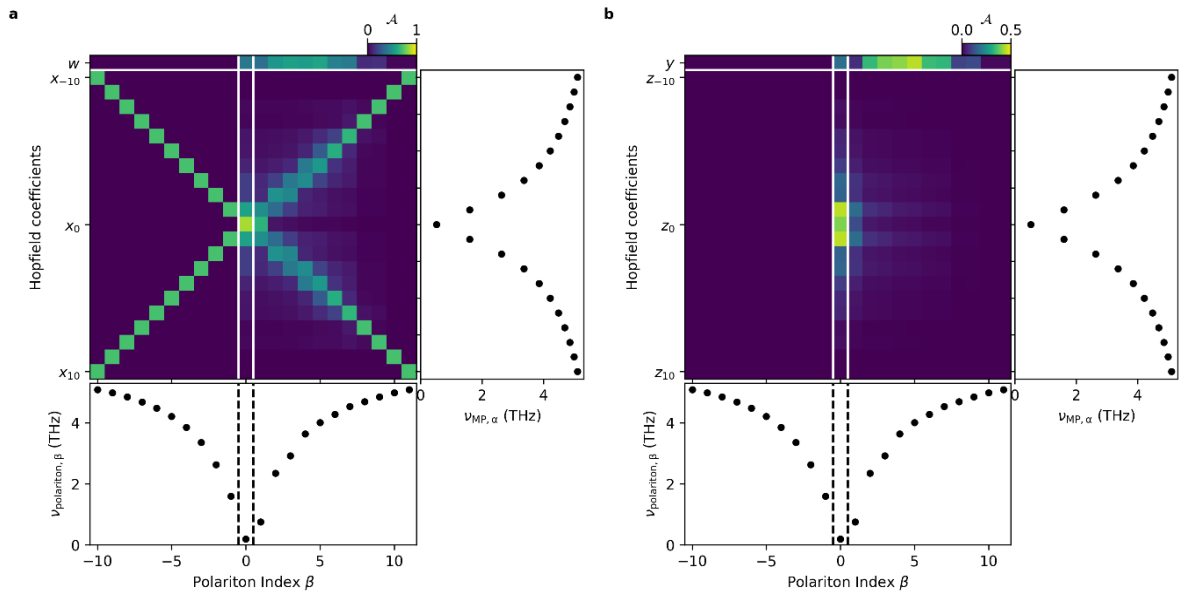

**Figure S5 | Hopfield coefficients for the 48-QW structure**. **a**, Plot of the absolute values of the complex-valued Hopfield coefficients for the lowest cavity mode, $j = 1$, and a cyclotron frequency of $\nu_c = 0.52$ THz. The right and bottom panels show the frequencies of the uncoupled magnetoplasmon and the coupled polariton modes, respectively. $w$: light mode, $x_{-10}$ to $x_{10}$: plasmon modes. Polariton index $\beta < 0$: uncoupled magnetoplasmon modes, $\beta = 0$: lower polariton, $\beta > 0$: upper polaritons. **b**, Absolute values of the anti-resonant Hopfield coefficients. $y$: light mode, $z_{-10}$ to $z_{10}$: plasmon modes.

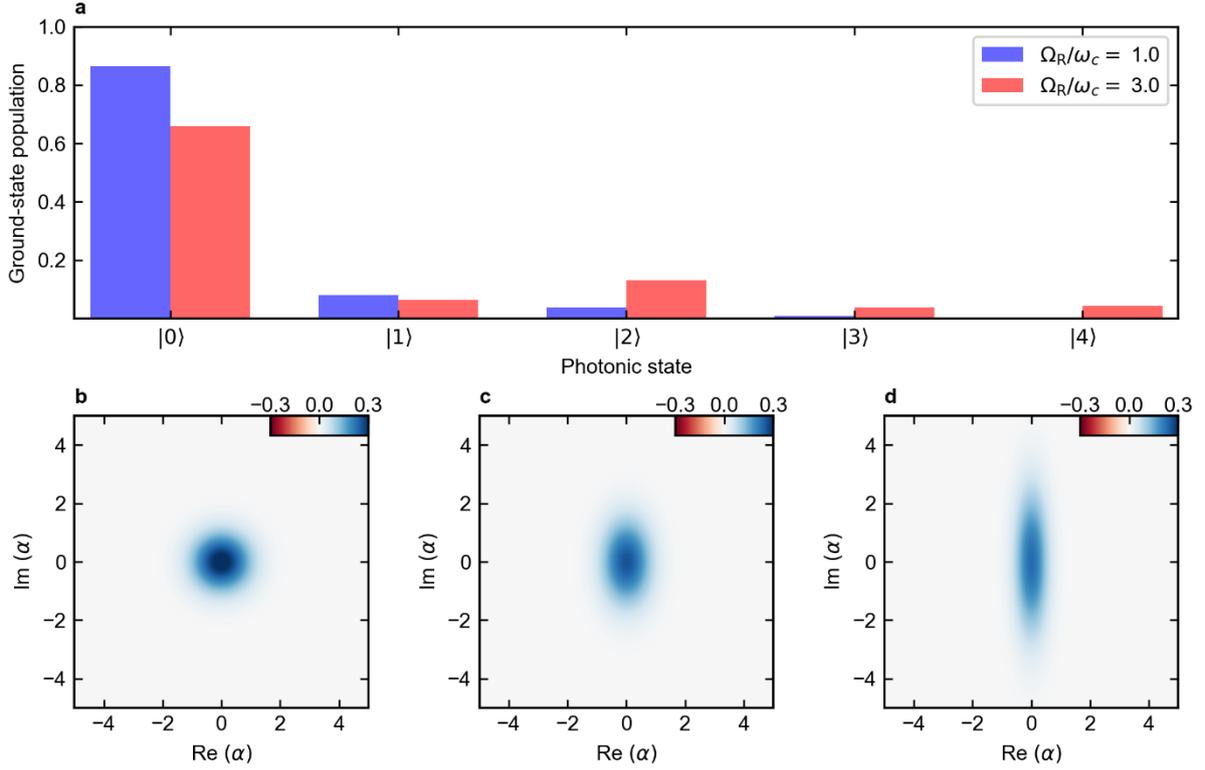

**Figure S6 | Virtual photon occupation and squeezing of the vacuum ground state.** The calculation considers a single cavity mode coupled to a single matter excitation, for clarity. **a**, Virtual photon population with a coupling strength of $\Omega_R/\omega_c = 1$ (blue bars) and $\Omega_R/\omega_c = 3$ (red bars). **b-d**, Wigner function of the photonic state for $\Omega_R/\omega_c = 0$ (**b**), $\Omega_R/\omega_c = 1$ (**c**) and $\Omega_R/\omega_c = 3$ (**d**).

Moreover, we investigate the virtual photon population $\langle N_1 \rangle$ as a function of the detuning of light and matter modes. To this end, we first calculate $\langle N_1 \rangle$ for a single cavity mode with a frequency of $\nu_{j=1} = 0.52$ THz which is coupled to a single electronic excitation with a variable frequency $\nu_c$, whereby we assume a coupling strength of $\Omega_{R,1}/\omega_1 = 2.83$ to approximate the situation of the structure with 48 QWs (Fig. S7a, dotted grey curve). Remarkably, even for a significant detuning of $\nu_c = 10 \times \nu_{j=1}$, the vacuum photon population remains above 50% of its maximum value found for $\nu_c = \nu_{j=1}$. The situation is even more favourable for the actual multi-mode setting of our structure, where tuning of the cyclotron resonance $\nu_c$ to values $\nu_c > \nu_{j=1}$ increases the detuning for some modes, while the detuning for others is reduced. As a result, the vacuum photon population (Fig. S7a, solid black curve) lies even slightly above the one for the single-mode scenario.

The situation changes considerably for moderate coupling strengths. We perform the calculation for a single pair of light and matter modes for $\Omega_{R,1}/\omega_1 = 0.1$. Here, a much stronger dependence of $\langle N_1 \rangle$ on the detuning is observed (Fig. S7b), illustrating that multi-mode coupling over multiple optical octaves crucially relies on significant light-matter coupling strengths.

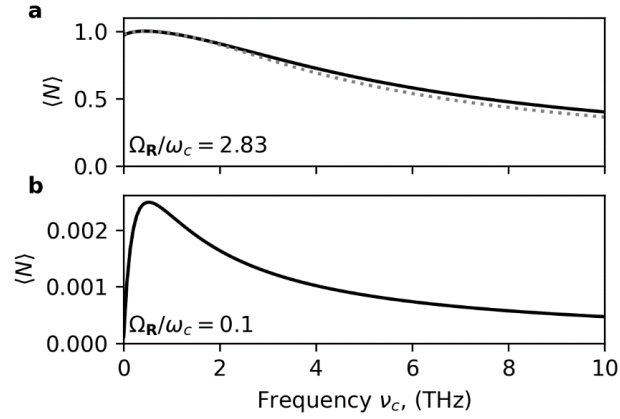

**Figure S7: Virtual photon population $\langle N \rangle$ as a function of detuning with the cyclotron frequency $\nu_c$** | **a,** A situation comparable to that of the structure featuring 48 QWs (black) with an equivalent single-mode coupling strength of $\Omega_{R,1}/\omega_1 = 2.83$, whereby only a single cavity and matter mode (grey dashed line) are considered with equivalent coupling strengths. **b,** For the single modes with a coupling strength of $\Omega_{R,1}/\omega_1 = 0.1$.

## 4. Subcycle time-domain quantum model

In order to compare the spectrum of coupled modes of our Hamiltonian to the spectra of the experimental data as well as the FEFD simulation, we derive the equations of motion for each operator using Heisenberg's equation of motion according to

$$\frac{d\hat{A}(t)}{dt} = -\frac{i}{\hbar}[\hat{A}(t), \hat{H}], \tag{14}$$

where $\hat{A}(t)$ is a bosonic operator representing the cavity or polariton modes. All couplings between modes result from expanding the commutator. Applying corresponding bosonic commutation relations, we derive a system of time-dependent differential equations which we subsequently treat by a mean-field approach, allowing us to introduce phenomenological dephasing rates and to include the driving field. For example, the dynamics of the cavity modes evolves according to

$$\dot{\alpha}_j = \langle \dot{\hat{a}}_j \rangle = -\frac{i}{\hbar}\langle[\hat{a}_j, \hat{H}]\rangle - \gamma_j \alpha_j + \mathcal{E}_{\text{THz}}(t),$$

where $\alpha_j = \langle \hat{a}_j \rangle$ is the mean-field expectation value of the cavity electric field, $\gamma_j$ is the damping rate, and $\mathcal{E}_{\text{THz}}(t)$ is the THz far field which drives the cavity. We model the complex frequency response of the metasurface by a superposition of cavity fields $\alpha_j$, each of which has a custom frequency $\nu_j$, relative amplitude contribution, phase and damping rate such that their response optimally represents the far-field response obtained from FEFD calculations. Implementing a total of five oscillators with the parameters given in table T1, the FEFD result is faithfully reproduced (Fig. S8).

Our time-domain quantum model allows for investigating the role of each optical mode for the transmission spectrum of the structure, by a switch-off analysis. More specifically, the contributions of the most important cavity modes $j = 1$ and $j = 2$ to the far-field, are separately displayed in Fig. S10.

| Mode index $j$ | Frequency $\nu_j$ (THz) | Damping rate $\gamma_j$ (THz) | Relative amplitude $A_j$ | Phase $\phi_j$ |
|---|---|---|---|---|
| 1 | 0.52 | 0.08 | 2.8 | 0 |
| 2 | 1.95 | 0.80 | 44.5 | $-0.14\,\pi$ |
| 3 | 3.75 | 0.12 | 0.8 | $-0.25\,\pi$ |
| 4 | 4.60 | 0.30 | 0.9 | $-0.6\,\pi$ |
| 5 | 6.00 | 0.30 | 7 | $-0.14\,\pi$ |

**Table T1** | Parameters for the cavity modes used to model the frequency response of the resonator structure.

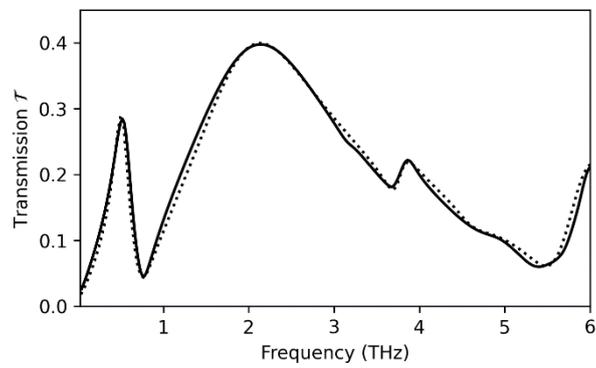

**Figure S8** | Far-field transmission of the bare metasurface calculated by the FEFD method (solid curve) and transmission obtained from the harmonic oscillator representation of our time domain simulations (dotted curve).

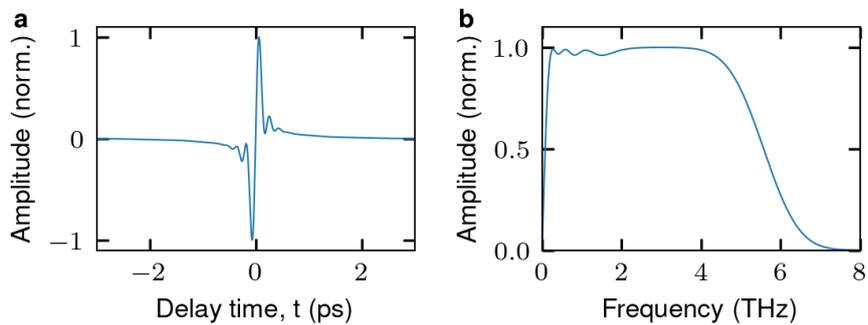

**Figure S9** | Characteristics of THz excitation used in the time-domain theory. **a**, Waveform and **b**, amplitude spectrum.

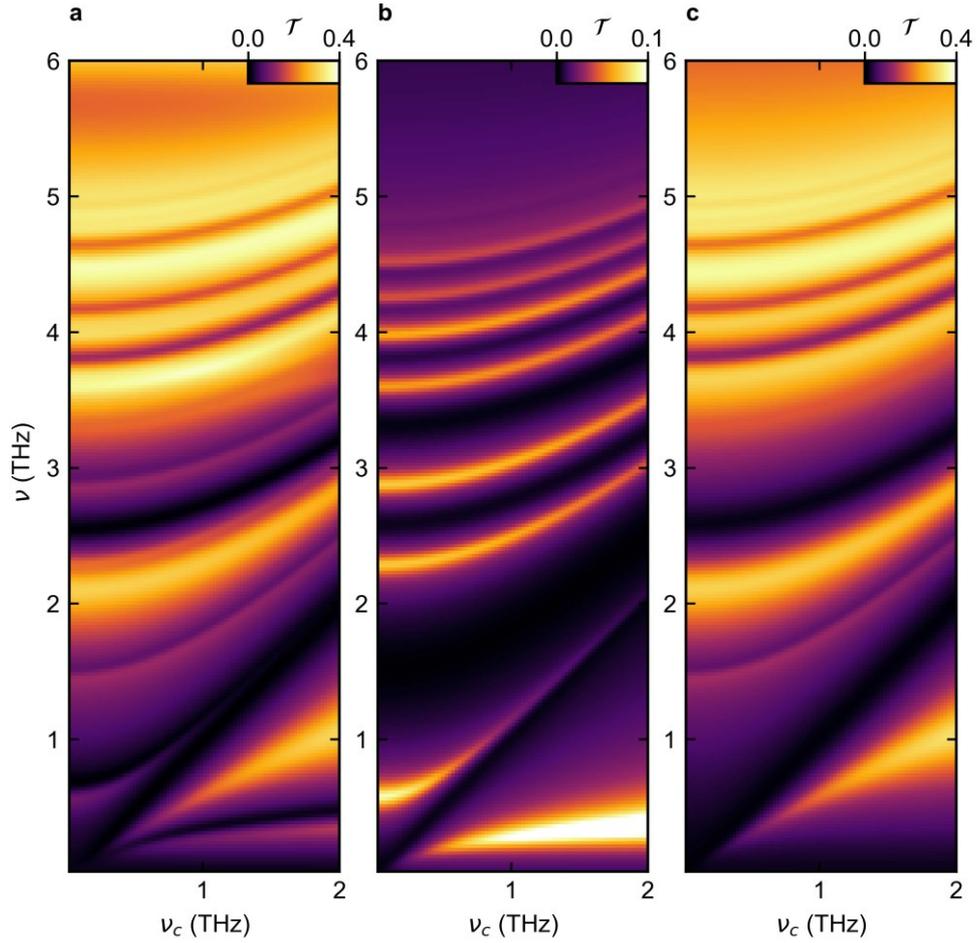

**Figure S10 | Switch-off analysis for the optical modes.** The time-domain simulation of the 48-QW structure includes **a,** all modes, **b**, only the first and **c**, only the second optical mode of the resonator structure.

As discussed in the main manuscript, our model allows us to access the microscopic polarization dynamics of the internal degrees of freedom of our multi-mode coupled structure. Owing to the extremely strong coupling strengths $\Omega_{R,j=1,\alpha}$, all individual MP modes significantly influence each other by coupling to the same cavity mode. This effect is illustrated in the amplitude spectra in Fig. S11. While the spectral weight of each MP mode is centered near its intrinsic resonance frequency, they all share a common set of frequencies where local maxima are observed – a characteristic signature of extremely strong mutual coupling of oscillators.

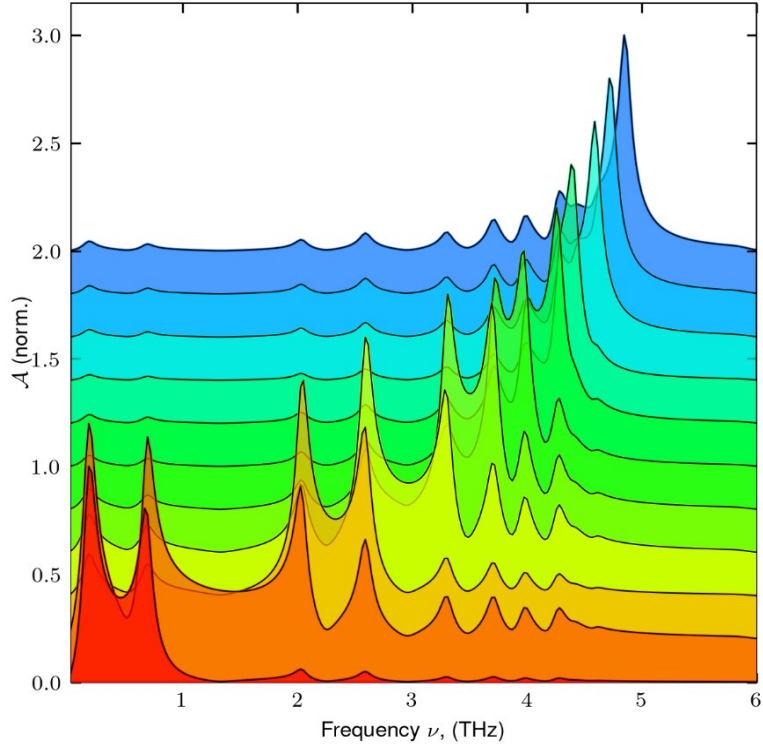

**Figure S11** | Calculated spectra of the expectation values of the polarization of the MP modes $0 \leq \alpha \leq 10$. The spectra are vertically offset by a value of 0.2, for clarity.

## 5. Parameter-free FEM model

In addition to our time-domain quantum model, we calculate the optical response of the bare resonator structure by numerical FEFD simulations following the concept of Ref. [6]. The calculation yields the response of the coupled structures including the spatially resolved near-field distribution as well as the far-field transmission, without any free fit parameters. Our formalism requires the three-dimensional geometry of the nanostructure including the generally anisotropic dielectric response, implemented as a tensor function $\underline{\epsilon}(\vec{r}, \omega_c, \omega)$ which depends on the position $\vec{r}$, the cyclotron frequency, $\omega_c$, and the frequency of the light field, $\omega$. Within the two-dimensional electron gas, $\underline{\epsilon}$ describes the gyrotropic nature of the cyclotron resonance in $x$ and $y$ direction owing to the $z$-polarized static magnetic bias field. In the $z$ direction, we employ the background dielectric constant, since the small thickness of the quantum wells leads to a plasma frequency far above the frequencies of interest for our structure. Additionally, we reduce the numerical complexity by modelling the quantum well stack including its barriers by an effective-medium approach with an effective dielectric tensor. The magnetically invariant background and substrate layers are implemented by the dielectric function $\epsilon_{\text{GaAs}}(\omega) = \epsilon_\infty \frac{\omega_{\text{LO}}^2 - \omega^2 + i\gamma_{\text{LO}}\omega}{\omega_{\text{TO}}^2 - \omega^2 + i\gamma_{\text{TO}}\omega}$, including the optical phonons of GaAs to improve the accuracy of the calculation especially at higher THz frequencies. Here we use $\epsilon_\infty = 10.87$ [10], $\omega_{\text{LO}}/2\pi = 8.839$ THz, $\omega_{\text{TO}}/2\pi = 8.124$ THz, $\gamma_{\text{LO}} = 0.0225$ THz and $\gamma_{\text{TO}} = 0.0255$ THz [11]. A single unit cell of the metasurface is implemented and periodically extended in $x$ and $y$ direction by periodic boundary conditions. The resulting far-field calculations predict experimental results across the entire spectral range with high accuracy (Fig. S12).

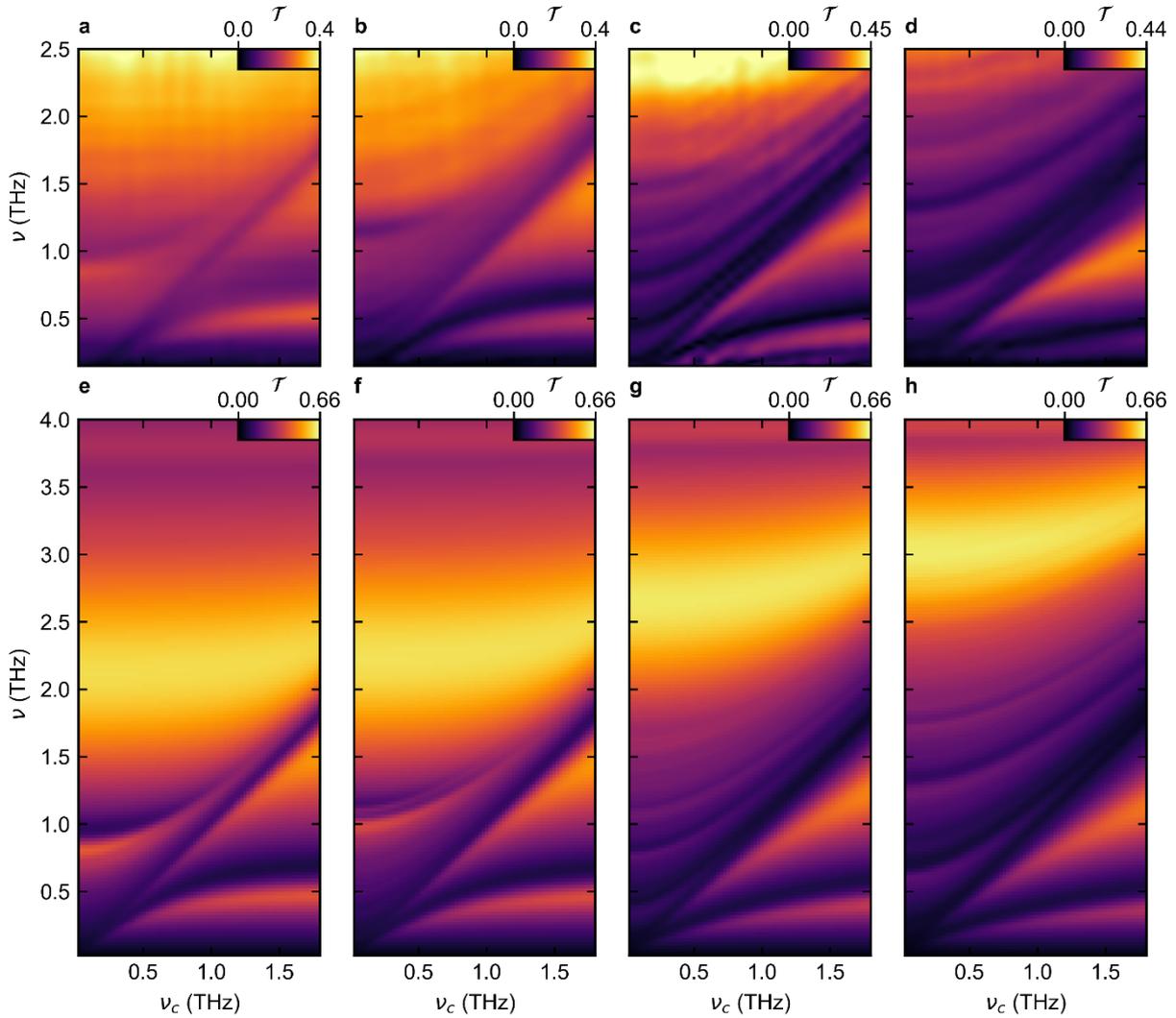

**Figure S12 | Comparison of experimental spectra and FEFD calculations. a,** Experimental THz magneto-transmission as a function of $\nu_c$ for the single-QW, **b,** 3-QW, **c, 6**-QW, and **d,** 12-QW structure. **e-h** Corresponding FEFD THz transmission data.

## 6. Scaling of the coupling strength

Following Hagenmüller et al. [12], the coupling strength of an ensemble of electronic oscillators coupled to a cavity mode scales with the square root of the total carrier density in the QW stack, $\Omega_R \propto \rho^{0.5}$. This dependency is perfectly reproduced for our first four structures containing 1, 3, 6 or 12 QWs. However, owing to limited penetration depth of the near-field into the QW stacks, we observe a deviation from this ideal scaling law for the structures with 24 and 48 QWs (Fig. S13). Nevertheless, the coupling strength still increases significantly even including this effect.

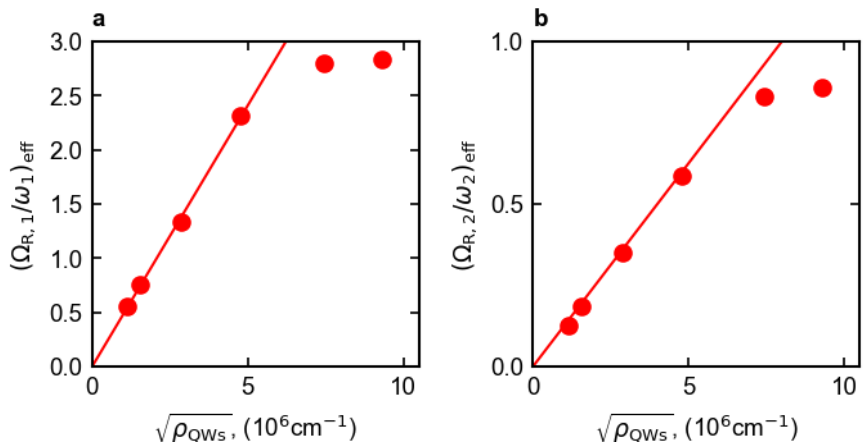

**Figure S13 | Scaling of the coupling strength with the number of electronic oscillators. a**, Equivalent coupling strength $\Omega_{R,1}/\omega_1$ for the first cavity mode as a function of the square root of the total charge carrier density, and linear fit of the data adjusted to the first four data points. **b**, Equivalent data for the second cavity mode.